\documentclass[prl,aps,twocolumn,showpacs,floatfix]{revtex4}
\usepackage{graphicx}
\usepackage{epsfig}
\usepackage{amsmath}
\usepackage{amssymb}
\usepackage{amsfonts}
\usepackage{mathptmx}
\usepackage{eucal}
\usepackage{bm}

\newcommand{\lam}{\lambda_\perp}
\newcommand{\Ic}{I_c^{\textit{GL}}}
\begin{document}

\title{Phase diagram of a current-carrying superconducting film in absence
of the magnetic field}
\author{E.V.Bezuglyi and I.V.Zolochevskii\footnote{e-mail:
zolochevskii@ilt.kharkov.ua}}
\affiliation{B.Verkin Institute for Low Temperature Physics and Engineering,
Kharkov 61103, Ukraine}

\begin{abstract}
We present the phase diagram for the current states of superconducting films,
based on the experimental investigation of the resistive transition induced
by transport current. We found that a rather narrow film with the width $w <
5\lam(T)$ ($\lam$ is the penetration depth of the magnetic field) never
enters the vortex state, but experiences direct transition from the purely
superconducting state to the resistive state with phase-slip centers as soon
as the current exceeds the Ginzburg-Landau critical current $\Ic$. The
Meissner current state of the films of intermediate width, $5\lam < w < 10
\lam$, transforms at $I > 0.8\Ic$ to the vortex resistive state which exists
within the current interval $0.8\Ic < I < I_m$, where the value $I_m$ of the
upper critical current is in a good agreement with the theory. The vortex
state of wide films, $w > 10\lam$, is realized within the current region
$I_c^{AL} < I < I_m$, where $I_c^{AL}$ is the transition point to the vortex
state calculated for the limiting case $w \gg \lam$. At $I>I_m$, the films
with the width $w > 5\lam (T)$ enter a vortex-free resistive state with
phase-slip lines.
\end{abstract}


\maketitle

\begin{figure}[tb]
\centerline{\includegraphics[width=8.5cm]{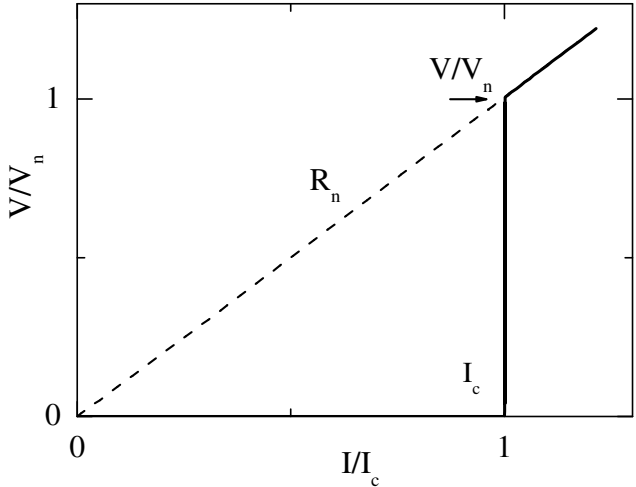}}
\caption{Current-voltage characteristic (IVC) of a narrow superconducting
channel according to the GL theory. Here $R_n$ is the channel resistance in
the normal state, and $V_n$ is the voltage jump at the point of the resistive
transition.} \label{fig1} \vspace{-5mm}
\end{figure}
\begin{figure}[bt]
\centerline{\includegraphics[width=8.5cm]{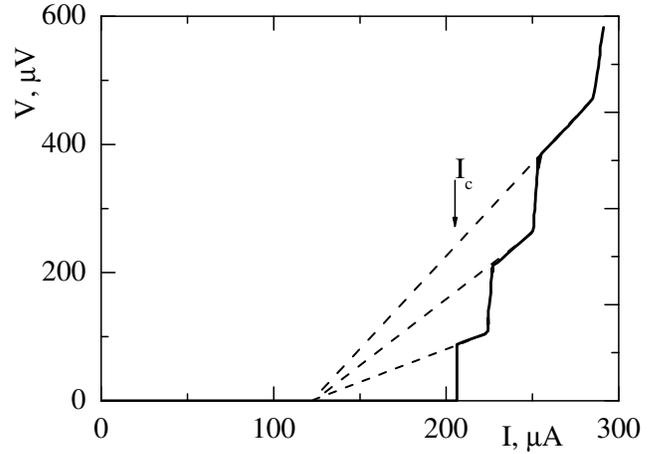}} \caption{Typical IVC of
a superconducting film channel Sn4 at the temperature $T/T_c =
0.98$.}\label{fig2}\vspace{-5mm}
\end{figure}

According to the Ginzburg-Landau (GL) theory \cite{1}, if the transport
current through a narrow superconducting film (superconducting channel)
exceeds the depairing current,
\begin{equation}
\label{eq1} I_c^{GL} (T)=\frac{c\Phi_0 w}{6\sqrt 3 \pi^2\xi (0)\lam
(0)}(1-T/T_c )^{3/2},
\end{equation}
the superconducting state of the channel is destroyed and transforms to the
normal state, as shown in Fig.\ref{fig1}. In (\ref{eq1}), $\Phi_0 $ is the
magnetic flux quantum, $w$ is the film width, $\lam(0)=2\lambda^2(0)/d$ is
the penetration depth of the magnetic field into the superconducting film,
$\xi (0)$ and $\lambda(0)$ are the coherence and London lengths,
respectively, at zero temperature, and $d$ is the film thickness. Later it
was found that the real scenario of the resistive phase transition of the
superconducting channel is more complex. Namely, as the transport current
exceeds $\Ic$, an inhomogeneous resistive state appears in the channel,
consisting of alternating superconducting and quasi-nor\-mal regions
\cite{2}; the latter are the specific dynamic formations known as phase-slip
centers (PSCs). The number of PSCs increases with the transport current, and
at $I>I_{cn} \gg \Ic$ the resistive state turns to the completely normal
state. The basic feature of the current-voltage characteristics (IVCs) of
superconducting channels in the resistive state are regular voltage steps
(Fig. \ref{fig2}), which were first observed in tin whiskers \cite{3,4} and
in narrow tin films \cite{5}. We note the following important peculiarities
of the step-like IVCs: the multiplicity of differential resistances of the
sloping IVC parts, the intersection of the continued sloping parts (dashed
lines) at the single point on the current axis, and the absence of hysteresis
which indicates the non-heating nature of the voltage steps.

\begin{figure}[th]
\centerline{\includegraphics[width=8.5cm]{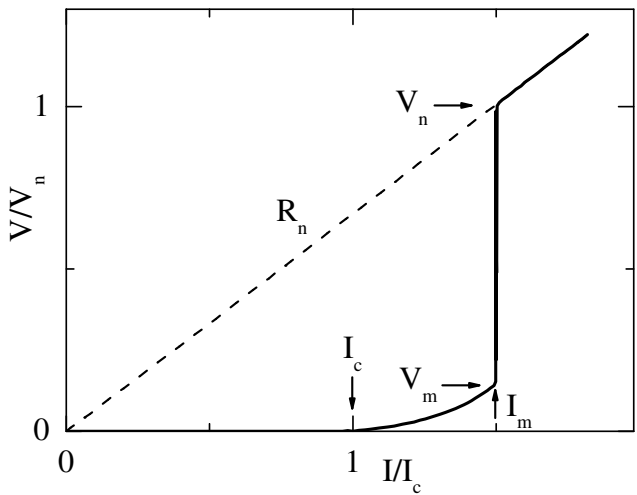}} \caption{Break-off-type
IVC of a wide superconducting film.}\label{fig3}\vspace{-3mm}
\end{figure}

To the present time, the resistive state of superconducting channels has been
rather well studied experimentally, and the theory of this state has been
universally recognized. In our opinion, this is not the case for the
resistive state of wide films, in spite of the fact that the study of
current-carrying states in wide films started much earlier than the
investigations of superconducting channels. In 1963, Tinkham \cite{6} first
involved the vortex conception for calculation of the critical fields of thin
films. It was found in \cite{7} that the motion of vortices, induced by the
magnetic field of the Earth or the transport current, plays the crucial role
in formation of the initial part of the IVCs. In next studies, the resistive
state of wide films was associated only with the vortex motion. This looked
quite natural because at that time, typical observed IVCs were similar to
that shown in Fig.\ref{fig3}, with abrupt transition from the vortex
resistive state to the normal state. We believe that in most cases, the
origin of such a break-off form of the IVC are imperfections of both the
experimental conditions and the samples. We note that in the experimental
studies of the resistive state, the heat compliance between the film and the
substrate plays an important role \cite{8}. If the choice of the pair
``film-substrate'' is not optimal, or the adhesion of the film to the
substrate is imperfect, the Joule overheating of the film in the resistive
state leads to the break-off IVCs shown in Figs.\ref{fig1} and \ref{fig3}.

\begin{figure}[th]
\centerline{\includegraphics[width=8.5cm]{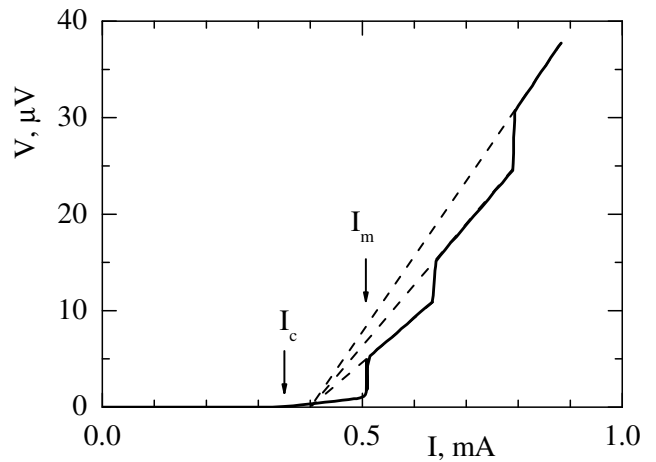}} \caption{Experimental
IVC of a wide film SnW13 at $T=0.99T_c$.} \label{fig4}\vspace{-3mm}
\end{figure}

In 1972, in study of wide tin films sputtered on the quartz substrate which
provides optimal heat compliance, the authors of \cite{9} observed not only a
typical vortex region of the IVC at small transport current, but also a
step-like structure at large current, which obtained no physical explanation
at that time. Relying on the recommendations given in \cite{8}, the authors
of \cite{10} fabricated the films having the IVCs similar to that shown in
Fig. \ref{fig4}. Then, using their knowledge about the phase-slip processes
in narrow channels, they associated the voltage steps in the IVCs of wide
films with the creation of phase-slip lines (PSLs). At the present time, it
seemed that such a form of the IVC for a wide film, consisting of the vortex
region and the step-like part due to the PSLs, is widely recognized; see,
e.g., \cite{9,10,11,12,13}. However, the break-off-type IVCs are still
frequently observed \cite{14,15,16,17,18}.

Let us first discuss the theoretical models of the vortex resistive state
\cite{20,21} which do not explain the whole form of the IVC shown in Fig.
\ref{fig4}, but give a rather good description of the vortex part of the IVC.
In wide thin superconducting films, the magnetic field of the transport
current gives rise to the creation of Pearl vortices at the film edges. The
motion of the vortices across the film leads to the occurrence of a voltage
along the film. On the basis of such a picture of the resistive vortex state,
the equation for the critical current was found by Aslamazov and Lempitskiy
(AL) \cite{20},
\begin{equation}
\label{eq2} I_c^{AL}(T) = 1.5 \Ic(0)[\pi \lam(0)/w]^{1/2}(1-T/T_c),
\end{equation}
using the condition of stability of the Meissner state with respect to an
infinitely small perturbation of the superconducting order parameter.
Physically, at $I=I_c^{AL}$, the edge current density approaches the GL
critical value, and the edge barrier for the vortex entry disappears. In
order to describe the vortex part of the IVC, AL studied the viscous motion
of the vortices in the film, using the hydrodynamic approximation which
assumes introduction of averaged macroscopic quantities: the density of
vortices and the averaged current density satisfying the macroscopic
equations which connect these quantities with the averaged electric field.
According to this theory, the evolution of the resistive vortex state looks
as follows. As the current grows, the vortex density increases, and the
current distribution across the film becomes more homogeneous. At a certain
current value,
\begin{equation}
\label{eq3} I_m(T) = C\,\Ic(T) \ln^{-1/2}[2w/\lam(T)] \gg I_c^{AL},
\end{equation}
the current density approaches its critical value not only at the film edges,
where the vortices are born, but also in the middle of the film
cross-section, where the vortices and anti-vortices annihilate. At this
moment the vortex state becomes unstable, although the distance between
vortices is still larger than the size of the vortex core, and the authors of
Ref. \onlinecite{20} assert that the film undergoes jump-like transition to
the normal state, as shown in Fig.\ref{fig3}. The experimental study of wide
films \cite{12} confirmed much of the statements of the AL theory, including
correctness of equations (\ref{eq2}) and (\ref{eq3}). Besides, these
investigations also resulted in considerable refinements of several points of
the theory. It turned out \cite{12} that the vortex resistivity occurs only
at large enough film width, $w>5\lam(T)$, and at $I>I_m$ the film undergoes
transition not to the normal state, but to a vortex-free state with PSLs
(Fig.\ref{fig4}). Such a picture of the resistive state of a wide film was
later recognized in \cite{22} by Lempitskiy. He predicted that if the
distances between PSLs are larger than the penetration depth for the electric
field into the superconductor, then the IVC of a wide film will be described
by known equations for a vortex-free narrow superconducting channel \cite{2}.
This result was also confirmed experimentally \cite{23,24}.

Another approach to the analysis of the vortex mechanism of resistivity was
used by Vodolazov and Peeters \cite{21} which performed numerical solution of
extended time-dependent Ginzburg-Landau equations \cite{25,26} for the vortex
motion in the superconducting plate, formally considering an infinitely thick
superconducting slab which reduces the problem to a two-dimensional one. In
this aspect, the problem becomes rather close to the problem of the vortex
state in a thin film, although the structure of vortices in these cases is
quite different: exponential decay of the magnetic field around the Abrikosov
vortices in a massive slab and slow power-like decay around the Pearl
vortices in a thin film. The results of \cite{21} generally confirm the
macroscopic analysis of the vortex state by AL \cite{20}: when the current
exceeds the critical value, at which the surface barrier for the vortex entry
is suppressed, the vortices and anti-vortices enter the slab and then, being
affected by the transport current, move to the middle of the sample, where
they annihilate. This process leads to a maximum of the current density in
the middle of the sample, in accordance with macroscopic calculations in
\cite{20}. At moderate values of the transport current, the vortex structure
of the sample is close to the triangle lattice. However, when the current
density in the middle of the sample approaches the depairing value, i.e.,
when the transport current approaches $I_m$, the triangle vortex lattice
turns into row-like vortex structure. The authors of \cite{21} interpret this
phenomenon as the creation of quasi-PSLs due to acceleration of the vortex --
anti-vortex annihilation and anomalously rapid motion of the vortices.
According to this theory, the PSLs represent the rows of rapidly moving
vortices which occur simultaneously along the whole length of the sample.
This contradicts the experimental data shown in Fig. \ref{fig4} which
indicate consequent appearance of the PSLs while the transport current
increases. For this reason, in order to confirm their conclusions, the
authors of \cite{21} refer only to the experiments \cite{14,15,16,17,18}, in
which the break-off-like IVCs were observed. In our opinion, these
experiments have mutual drawback: the substrates do not provide a good heat
removal from the films. This is indicated by hysteresis observed, e.g., in
\cite{16}, or by low quality of the films with numerous pinning centers and
bad adhesion with the substrate, as in \cite{15}. We believe that for these
reasons, the PSLs occur in such films in an avalanche, which is accompanied
by a break-off in the IVC. Our point of view is supported by the results of
investigation of the resistive state of wide films by the laser scanning
microscope \cite{19}. Due to specifics of this method, the film is overheated
by the laser irradiation, which gives rise to the break-off behavior of the
IVC (Fig. \ref{fig3}). However, the visualization of the resistive state by
this microscope shows that at $I>I_m$ the number of PSLs increases smoothly
with the transport current, starting from the single PSL at $I=I_m$.

\begin{table}[htbp]
\begin{center}
\begin{tabular}{|l|l|l|l|l|l|l|l|l|}
\hline Sample& $L$, $\mu $m& $w$,$\mu $m& $d$, nm& $R_{4.2}$, $\Omega $&
$R^{\square}$, $\Omega $& $T_{c}$, K& $l_{i}$, nm&
$R_{300}$, $\Omega $ \\
\hline Sn4& 30& 1& 199& 1.45& 0.048& 3.783& 131&
21.50 \\
\hline SnW9& 95& 17& 159& 0.319& 0.057& 3.825& 138&
4.900 \\
\hline SnW10& 88& 7& 181& 0.487& 0.040& 3.809& 169&
9.156 \\
\hline SnW13& 90& 18& 332& 0.038& 0.008& 3.836& 466&
1.880 \\
\hline
\end{tabular}
\label{tab1}
\end{center}
\caption{Parameters of the film samples. Here $L$, $w$ and $d$ are the
length, the width and the thickness of a sample; $l_{i}$ is the electron mean
free path.}
\end{table}

Now we proceed to the results of our experimental studies. The samples whose
characteristics are given in the Table were fabricated by using an original
technique which provides minimization of bulk defects and results in perfect,
almost specular, film edges. All such samples show full GL critical current
in a near vicinity of $T_c$, where the condition of narrow channel regime $w
< \lam(T)$ is satisfied; actually, this criterion was used for the selection
of samples for next experiments in the resistive state. Special attention was
devoted to the quality of the substrate - optically polished crystalline
quartz, which seems to be the better material for the heat removal from tin
films.

\begin{figure}[tb]
\centerline{\includegraphics[width=8.5cm]{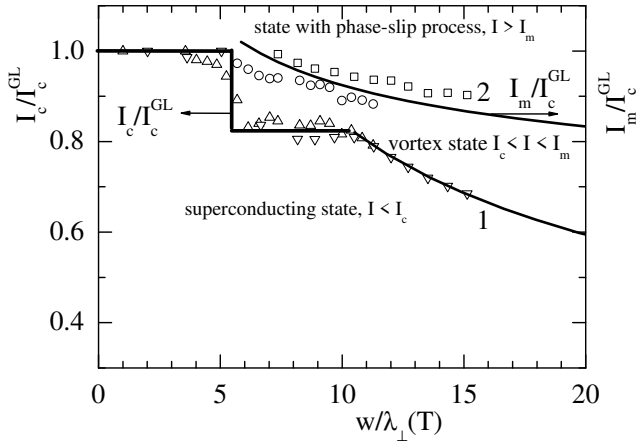}} \caption{Diagram of the
current-carrying states of wide superconducting films in the dependence on
the reduced film width and the reduced magnitude of the transport current.
Different states are separated by the dependencies of the reduced critical
current $I_c/\Ic$ for the samples SnW9 ($\triangledown$), SnW10 ($\triangle$)
and of the reduced current of the vortex instability $I_m/\Ic$ for the
samples SnW9 ($\square$), SnW10 ($\bigcirc$) on the quantity $w/\lam$. The
smooth part of the curve 1 represents the AL theoretical dependence of
$I_c^{AL}/\Ic$ calculated by (\ref{eq1}) and (\ref{eq2}); the curve 2 is the
AL theoretical dependence of $I_m/\Ic$ calculated by (\ref{eq1}) and
(\ref{eq3}) with $C=1.6$.}\label{fig5}\vspace{-3mm}
\end{figure}

As follows from the AL theory \cite{20} [see (\ref{eq2}) and (\ref{eq3})],
the reduced critical currents $I_c^{AL}/\Ic$ and $I_m/\Ic$ for wide films,
$w/\lam \gg 1$, must be universal functions of the basic parameter of the
theory, $w/\lam (T)$. In other words, the dependencies of the reduced
critical currents on this parameter should be not affected either by the
geometry of the films, or by their material properties. Our experimental
results completely confirm such universality and, moreover, extend it over
the small values of $w/\lam (T)$. Examples of the experimental dependencies
of the reduced critical currents are presented in Fig. \ref{fig5}. To avoid
overloading of the figure, we plot the results obtained on a single pair of
essentially different films; other samples demonstrate quite similar
behavior. The triangles (direct and inverted) correspond to the upper
boundary of the purely superconducting state. In the regime of narrow film
(superconducting channel), $w/\lam (T) < 5$, the sample is completely
superconducting until the transport current reaches the GL depairing current;
at $I \geq \Ic$ the film undergoes transition to the resistive state with
one-dimensional PSCs. Correspondingly, this part of the dependence of
$I_c/\Ic$can be approximated by horizontal straight line. When the ratio
$w/\lam (T)$ exceeds 5, an unusual phenomenon is observed: the value of the
critical current sharply falls to $0.8\Ic$ (vertical approximating line) and
then holds this value until $w/\lam (T)$ reaches 10 (second horizontal line).
As soon as the transport current exceeds $0.8\Ic$, the film enters the vortex
resistive state, in accordance with general conclusions of the theory.
However, such behavior of $I_c$ is inconsistent with the AL theory, even with
its generalized version valid for the case of arbitrary value of $w/\lam (T)$
and developed by us. Formal consideration of this problem leads to the
conclusion that at $5 < w/\lam (T) < 10$ the vortices can overcome the edge
barrier when the edge current density approaches the value $j_c \sim
(1-T/T_c)^2$ much smaller than the GL critical density $j_c^{GL}$, but the
physical mechanism of such anomalous penetration of vortices into the film is
unclear. For very wide films with the transversal size $w > 10\lam$, the
experimental data excellently agree with the AL theory (curve 1 in Fig.
\ref{fig5}). The experimental values of the upper boundary of the stability
of the vortex state, $I_m/\Ic(T)$, are plotted in Fig. \ref{fig5} for two
different films by circles and squares. As it is obvious from Fig.
\ref{fig5}, these data well correlate with the result (\ref{eq3}) of the AL
theory with the fitting parameter $C=1.6$ (curve 2).

The set of lines in Fig.5 can be considered as critical lines at the phase
diagram for superconducting films. These lines divide the phase plane
``reduced current $I/\Ic$ -- reduced film width $w/\lam$'' into three
regions. The lower region, $I<I_c$, corresponds to the completely
superconducting state: the homogeneous current state in narrow films, $w <
5\lam$, or the Meissner state in wide films, $w > 5\lam$. For the latter
case, there exists the intermediate vortex resistive region, $I_c<I<I_m$,
where $I_c = 0.8\Ic$ at $5\lam < w< 10\lam$, and $I_c = I_c^{AL}$ for $w
>10\lam $. Then, at $I>I_m$ the wide film ($w>5\lam$) enters the resistive
state with PSLs, while the narrow film ($w<5\lam$) exhibits direct transition
to the resistive state with PSCs at $I>\Ic$.

In conclusion, we develop the phase diagram for the current states of
superconducting films, based on the experimental investigation of the
resistive transitions induced by the transport current in the films. We found
that a narrow film with the width $w<5\lam$ never enters the vortex state,
but experiences direct transition from the purely superconducting state to
the resistive state with PSCs as soon as the current exceeds the critical GL
current. The Meissner current state of the films of intermediate width,
$5\lam < w< 10\lam$, transforms at $I>0.8\Ic$ to the vortex resistive state
which exists within the current interval $0.8\Ic < I < I_m$, where the value
of the upper critical current $I_m$ well agrees with the theory \cite{20}.
The vortex state of wide films, $w>10\lam$, is realized within the current
region $I_c^{AL}<I<I_m$, where $I_c^{AL}$ is the transition point to the
vortex state calculated in \cite{20} for the limiting case $w \gg \lam$. At
$I>I_m$, the films with the width $w>5\lam$ enter a vortex-free resistive
state with phase-slip lines.

\end{document}